\documentclass[conference]{IEEEtran}
\usepackage{cite}
\usepackage{threeparttable}
\usepackage{graphicx}
\usepackage{picinpar}
\usepackage[cmex10]{amsmath}
\usepackage{amsmath,amsfonts,amssymb}
\usepackage[ruled,lined,boxed]{algorithm2e}
\usepackage{algorithmic}
\usepackage{subfig}
\usepackage{threeparttable}
\usepackage{caption}
\usepackage{booktabs}
\usepackage{multirow}
\usepackage{etoolbox}
\usepackage{makecell}
\usepackage{amssymb}
\usepackage{mathrsfs}
\usepackage{color}
\definecolor{purple}{RGB}{160,32,240}
\definecolor{darkred}{RGB}{255,0,255}
\usepackage{stfloats}
\usepackage{bm}
\begin{document}
	\IEEEoverridecommandlockouts
	\newtheorem{lemma}{Lemma}
	\newtheorem{corol}{Corollary}
	\newtheorem{theorem}{Theorem}
	\newtheorem{proposition}{Proposition}
	\newtheorem{definition}{Definition}
	\newcommand{\e}{\begin{equation}}
	\newcommand{\ee}{\end{equation}}
	\newcommand{\eqn}{\begin{eqnarray}}
	\newcommand{\eeqn}{\end{eqnarray}}
	
	\title{Three-Dimensional Trajectory Design for Multi-User MISO UAV Communications: A Deep Reinforcement Learning Approach}
	
	\author{Yang Wang, Zhen Gao
		
		\thanks{The work was supported by the NSFC under Grants 62071044 and 61827901, the BJNSF under Grant L182024.}
		\thanks{Y. Wang, Z. Gao are with the School of Information and Electronics, Beijing Institute of Technology, Beijing 100081, China (e-mail: gaozhen16@bit.edu.cn).}}
	
	\maketitle
	
	\begin{abstract}
		In this paper, we investigate a multi-user downlink multiple-input single-output (MISO) unmanned aerial vehicle (UAV) communication system, where a multi-antenna UAV is employed to serve multiple ground terminals. Unlike existing approaches focus only on a simplified two-dimensional scenario, this paper considers a three-dimensional (3D) urban environment, where the UAV's 3D trajectory is designed to minimize data transmission completion time subject to practical throughput and flight movement constraints. Specifically, we propose a deep reinforcement learning (DRL)-based trajectory design for completion time minimization (DRL-TDCTM), which is developed from a deep deterministic policy gradient algorithm. In particular, to represent the state information of UAV and environment, we set an additional information, i.e., the merged pheromone, as a reference of reward which facilitates the algorithm design. By interacting with the external environment in the corresponding Markov decision process, the proposed algorithm can continuously and adaptively learn how to adjust the UAV's movement strategy. Finally, simulation results show the superiority of the proposed DRL-TDCTM algorithm over the conventional baseline methods.
	\end{abstract}
	
	
	\IEEEpeerreviewmaketitle
	
	\section{Introduction}
	
	\IEEEPARstart{T}{}he unmanned aerial vehicle (UAV)-assisted communication paradigm is expected to play a pivotal role in the next-generation wireless communication systems, which promise to provide ubiquitous connectivity with broader and deeper coverage\cite{summary}. Particularly, using UAVs as aerial mobile base stations (BSs) to transmit data for distributed ground terminals (GTs) is anticipated to be a promising technology for realizing green communications\cite{UAV}. Compared to terrestrial BS-based communication systems, the UAV-based aerial BS system has salient attributes, such as a high probability in establishing strong line-of-sight (LoS) channels to improve coverage, a flexible deployment and fast response for unexpected or limited-duration missions, and a dynamic three-dimensional (3D) placement and movement for improving spectral and energy efficiency, etc\cite{UAV_sum}.
	
	Due to the high mobility, UAVs can move towards potential GTs and establish reliable connections with a low power consumption. Thus, the UAVs' trajectory design is essential for UAV-assisted communication systems. To date, there have been several related work investigating the trajectory design with various optimization targets, such as throughput, energy-efficiency, and flight time\cite{multi-Rate,multi-efficient,time}. In \cite{multi-Rate}, the authors considered to jointly optimize GTs' transmission scheduling, power allocations, as well as the multi-antenna UAV's two-dimensional (2D) trajectory for maximizing the minimum sum-rate in uplink communication. Besides, to minimize the total power consumption in multi-user multiple-input single-output (MISO) communication systems, authors in \cite{multi-efficient} jointly optimized the 2D trajectory and the transmit beamforming vector of the UAV. Also, in\cite{time}, the authors designed the UAV’s flight trajectory for minimizing the UAV cruising time for data transmission, so as to achieve the throughput, energy and delay requirements.
	
	However, the above UAV trajectory designs based on conventional optimization solutions have some critical limitations. First, formulating an optimization problem requires an accurate and tractable radio propagation model, which is often difficult to be obtained. Second, optimization-based design also requires the perfect channel state information (CSI), which is tough to acquire in practice. At last, most optimization problems in modern communication systems are highly non-convex and difficult to be efficiently solved.
	
	Considering these challenges, there have been several works leveraging deep reinforcement learning (DRL)\cite{Bible} for UAV-assisted communications. Specifically, in \cite{efficient_DRL1, efficient_DRL3}, the authors proposed a DRL-based UAV control method for maximizing the energy efficiency, data transmission, and fair communication coverage in mobile crowd sensing systems. In \cite{CCU_RL}, to minimize the weighted sum of the mission completion time and the expected communication outage duration, the authors focused on optimizing a UAV trajectory with the assistance of DRL. Nevertheless, existing DRL-based approaches usually assume a simplified channel model\cite{efficient_DRL1,efficient_DRL3}, or single service target scenario\cite{CCU_RL}, which may lead to a model mismatch with an unavoidable performance loss for practical urban scenario.

	To overcome the limitations above, this paper considers a 3D trajectory design for completion time minimization (TDCTM) in a multi-user downlink MISO UAV communication system. Specifically, the UAV with multi-antenna is employed to serve multiple GTs distributed in a 3D urban scenario. For such scenario, to cope with the continuous control problem with an infinite action space, we propose a DRL-based TDCTM (DRL-TDCTM) algorithm, which is conceived based on an actor-critic algorithm, called 	 \begin{figure}[htbp]
		\centering{\includegraphics[width=\columnwidth,keepaspectratio]{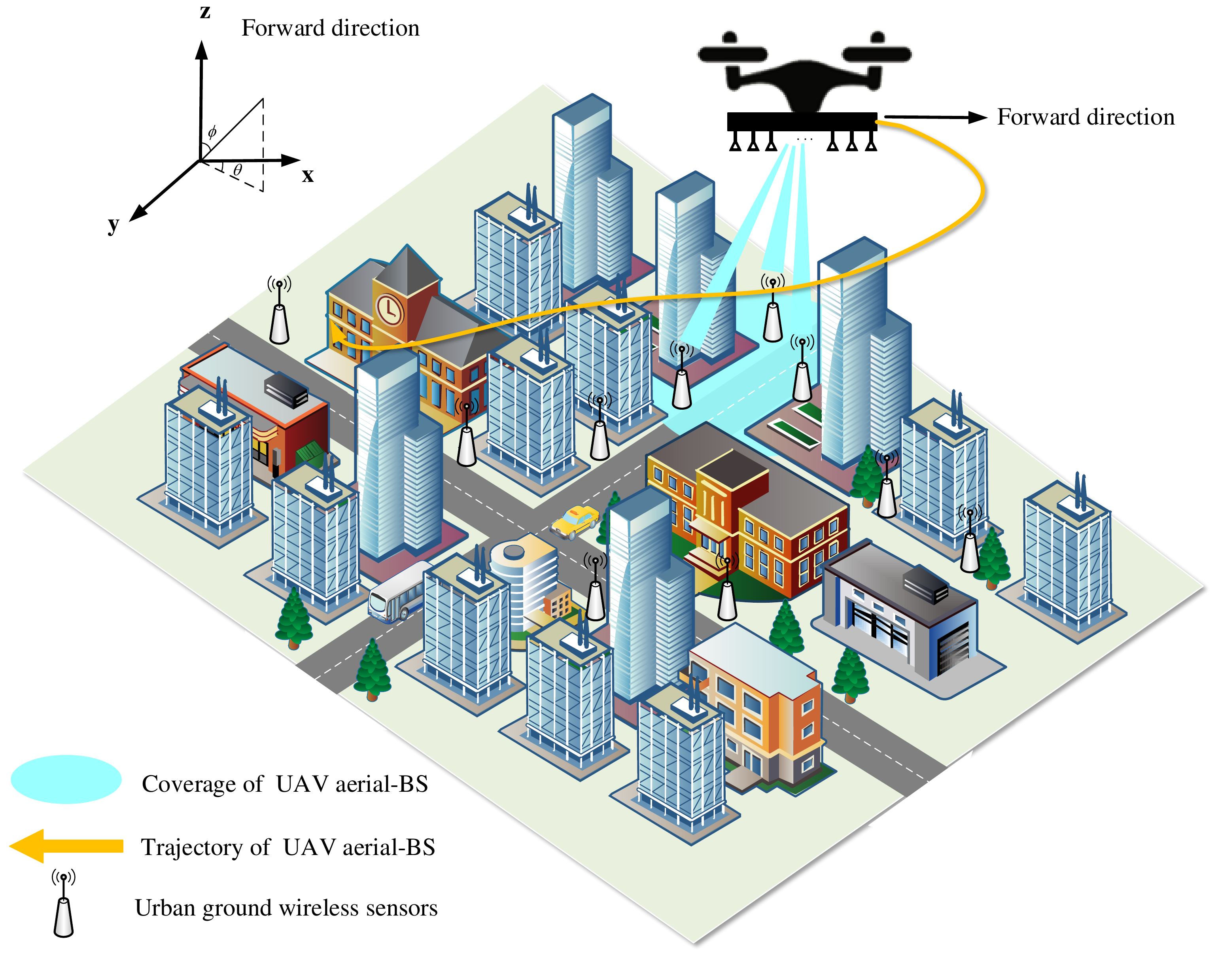}}
		\captionsetup{font={footnotesize}, name={Fig.},labelsep=period}
		\caption{Multi-antenna UAV-assisted MISO communication system.}
		\label{fig:scenario}\vspace{-3mm}
	\end{figure}deep deterministic policy gradient (DDPG)\cite{DDPG}. Besides, inspired by ant colony algorithm\cite{ant}, we set up an additional information, i.e., the merged pheromone, which is adopted as a input of reward function. Finally, simulation results verify the superiority of the proposed algorithm over the conventional baseline methods.
	
	\section{System Model and Problem Formulation}\label{II}
	\subsection{System Model}
	As shown in Fig. \ref{fig:scenario}, we consider a multiuser downlink MISO UAV communication system, where a UAV equipped with an $N_t$-element uniform linear array (ULA) is dispatched to serve a large number of single-antenna static GTs. We assume that $K$ GTs are randomly distributed in a given geographical region of $D\times D$ m$^2$ and the set of GTs is denoted by $\mathcal{K}=\left\{1,\cdots,K\right\}$. The positions of the $k$-th GT and the UAV are denoted by $\boldsymbol{w}_k=[\bar{x}_k,\bar{y}_k,0]\in\mathbb{R}^{3}$ and $\boldsymbol{q}(t)=[x_{t},y_{t},z_{t}]\in\mathbb{R}^{3}$, $0\le t\le T$, respectively, where $(\bar{x}_k,\bar{y}_k)$ denotes the horizontal coordinate of the $k$-th GT, $(x_{t},y_{t},z_{t})$ denotes the 3D Cartesian coordinate of the UAV, and $T$ is the mission execution duration.
	
	Compared with \cite{multi-Rate,multi-efficient,time}, we consider a more practical ground-air (G2A) channel model\cite{CCU_RL}, which can be characterized by large-scale fading and small-scale fading, and both of them are calculated based on a simulated 3D map by taking into account the existence of buildings as propagation scatterers. Specifically, the location and height of the buildings are generated according to a statistical model\cite{ITU}. In this model, there are three parameters to characterize an urban environment, including the ratio of land area covered by buildings to the total land area $\alpha$, the average number of buildings per square kilometer $\beta$, and the building height that can be modeled as a Rayleigh distribution with mean value $\lambda$.

	Given a specific area with the simulated building location and height, we can accurately determine whether there is a LoS link between the UAV and the $k$-th GT by checking whether the line connecting them is blocked by any building. Thus, the large-scale fading of the G2A channel associated with the $k$-th GT can be expressed as \cite{LAP}
	\begin{equation} {\rm PL}_{k}(t)=\begin{cases}L_{k}^{\rm{FS}}(t)+\eta_{\rm LoS},\\
	L_{k}^{\rm{FS}}(t)+\eta_{ \rm NLoS},
	\end{cases}
	\label{Eq1}
	\end{equation}
	where ${L_{k}^{\rm{FS}}}(t)=20\log d_{k}(t)+20\log f_c+20\log \left(\frac{4\pi}{c}\right)$ represents the free space pathloss between the UAV and the $k$-th GT, $d_k(t)=\left \| \boldsymbol{q}(t)-\boldsymbol{w}_k \right \|$ denotes the distance from the UAV to the $k$-th GT, $f_c$ denotes the carrier frequency, and $c$ represents the velocity of light. Besides, $\eta_{\rm LoS}$ and $\eta_{\rm NLoS}$ represent the propagation loss of the LoS and NLoS links, respectively\footnote{The above pathloss expressions are all in dB.}. Considering a MISO UAV communication system,  the baseband equivalent complex channel between the UAV and the $k$-th GT can be modelled as
	\begin{equation}
	\boldsymbol{h}_{k}(t)= 10^{-{\rm PL}_{k}(t)/20} {\boldsymbol{g}}_{k}(t),
	\end{equation}
	where ${\boldsymbol{g}}_{k}(t)$ denotes the small-scaling fading, which is modelled as the Rician fading with
	\begin{equation}
	\boldsymbol{g}_{k}(t)= \sqrt{\frac{G}{G+1}}\bar{\boldsymbol{g}}_k(t)+\sqrt{\frac{1}{G+1}}\tilde{\boldsymbol{g}}_k(t),
	\end{equation}
	where $G$ is the Rician factor, $\bar{\boldsymbol{g}}_k(t)$ is the steering vector function defined as
	\begin{equation}
	\bar{\boldsymbol{g}}_k\left(t\right)=\left[1,e^{j\pi\bar{\theta}_k},e^{j\pi 2\bar{\theta}_k},\cdots,e^{j\pi\left(N-1\right)\bar{\theta}_k}\right]^{\rm T},
	\end{equation}  
	where $\bar{\theta}_k$ represents the phase of the LoS path between the UAV and the $k$-th GT, and $\tilde{\boldsymbol{g}}_k(t)\sim\mathcal{CN}\left(\boldsymbol{0},\boldsymbol{{\rm I}}_{N_t}\right)$ denotes the Rayleigh fading channel component. As shown in Fig. \ref{fig:scenario}, the ULA of the UAV always maintains the forward direction vector $\left[1,0,0\right]$. Thus, the phase of the direct link can be expressed as $\bar{\theta}_k=\left(\bar{x}_k-x_t\right)/{d_k\left(t\right)}$. Furthermore, the Doppler effect caused by the UAV mobility is assumed to be well estimated and then compensated at the receiver\cite{Doppler}.

	\subsection{Problem Formulation}\label{A}
	To make the UAV's trajectory optimization problem tractable, the continuous time domain is discretized into $N$ time steps with unequal duration length $\delta_n$, $n\in\{0,1,\ldots,N\}$, and the data transmission task is performed within a series of time steps, i.e., $\{\delta_0,\delta_1,\ldots,\delta_N\}$. In addition, we consider that each time step consists of two parts, i.e., $\delta_n=\delta_{\rm ft}+\delta_{{\rm ht},n}$, where $\delta_{\rm ft}$ is the fixed flight time and $\delta_{{\rm ht},n}$ is the hovering time for data transmission. If there is no active GT in the current time step, the UAV would skip hovering and directly execute the next time step, i.e., $\delta_{{\rm ht},n}=0$ s. During each time step, the UAV's moving strategy can be expressed as
	\begin{align}
	x_{n+1}&= x_{n}+m_n\sin\left(\phi_n\right)\cos\left(\theta_{n}\right),\\
	y_{n+1}&= y_{n}+m_n\sin\left(\phi_n\right)\sin(\theta_{n}),\\
	z_{n+1}&= z_{n}+m_n\cos\left(\phi_n\right),
	\end{align}
	where $m_n=\delta_{\rm ft}{\upsilon_n}$ represents the moving distance of the UAV, $\upsilon_n\in[0,\upsilon_{\rm max}]$ denotes the average flight speed, $\upsilon_{\rm max}$ denotes the maximum cruising speed, $\phi_{n}\in [0,\pi] $ denotes the pitch angle of the UAV from the positive $z$-axis, and $\theta_{n}\in (0,2\pi] $ denotes the horizontal direction of the UAV in the $xy$-plane with respect to the $x$-axis. 
	
	Moreover, we consider a downlink communication system with three main steps: First, the UAV only activates a single-antenna for broadcast service to wake up the GTs which satisfy the communication requirement. Then, the active GTs will send control signals to the UAV through the uplink channels, and the UAV aerial BS will detect the active GTs and estimate the corresponding channels. Last, the downlink precoding is performed on the UAV according to the channel reciprocity of TDD system for downlink MISO data transmission service. Due to the assumption that active device detection and channel estimation can be addressed well\cite{active_DE}, we only pay attention to the first and last steps\footnote{Studying joint trajectory design, active device detection, channel estimation, and data transmission for fixed-wing UAVs is an interesting topic for future work.}.

	Thus, the channel gain from the UAV to the $k$-th GT during the broadcast stage in the $n$-th time step can be expressed as
	\begin{equation}
	h_{k,n}^1= 10^{-{\rm PL}_{k,n}/20} {g}_{k,n}.
	\end{equation}
	
	We assume that only when GT is awakened by the UAV, it can feed back its status to the UAV by the uplink channel, otherwise it continues to stay in the silent mode for energy saving. In the $n$-th time step, if the $k$-th GT is waken up, the corresponding signal-to-noise ratio (SNR) between the $k$-th GT and the UAV can be expressed as
	\begin{equation}
	\rho_{k,n}^1=\frac{P{\left\vert h_{k,n}^1 \right\vert}^2}{\sigma^2},
	\end{equation}
	where $P$ is the transmitter power during the broadcast stage and $\sigma^2$ represents the power of the additive white Gaussian noise (AWGN) at the ground receiver. For the downlink data transmission service associated with the $k$-th GT, we set a pre-defined SNR threshold $\rho_{\rm th}$, and the $k$-th GT can be awakened and served by the UAV if and only if $\rho_{k,n}^1 \ge \rho_{\rm th}$. Therefore, we define a binary variable $b_{k,n}\in\left\{0,1\right\}$ to indicate whether the $k$-th GT can satisfy the SNR requirement by the UAV in the $n$-th time step. Due to the assumption that each GT can only be served at most once in one realization, we define the following indicator function of the $k$-th GT as 
	\begin{equation}
	\tilde{b}_{k,n}=\begin{cases}1,\quad {\rm if}\, b_{k,n}=1, {\rm and}\, c_{k,n}=0, \\
	0,\quad {\rm otherwise},
	\end{cases}
	\label{eq-indict2}
	\end{equation}
	where $c_{k,n}\in\left\{0,1\right\}$ is a binary variable to indicate whether the $k$-th GT has been served by the UAV. Thus, we define the serving flag $c_{k,n}$ as
	\begin{equation}
	c_{k,n/0}=\min\left\{\sum_{i=0}^n \tilde{b}_{k,i}, 1\right\},c_{k,0}=0\label{eq-indict3}
	\end{equation}
	where if $c_{k,n}=1$, the $k$-th GT has been served during the mission; otherwise, the $k$-th GT has not been served.
	
	Define $\mathcal{K}_n=\left\{k\in \mathcal{K}:\tilde{b}_{k,n}=1 \right\}$ as the set of the active GTs in the transmission stage of the $n$-th time step and $K_n=\left\vert\mathcal{K}_n\right\vert$. When $K_n\ne0$, the corresponding channel vectors can be expressed as
	\begin{equation}
	\boldsymbol{h}_{k,n}^2= 10^{-{\rm PL}_{k,n}/20} {\boldsymbol{g}}_{k,n},k\in\mathcal{K}_n.
	\end{equation}
	
	Therefore, the corresponding channel matrix is defined by $\boldsymbol{H}_n=\left[\boldsymbol{h}_{1,n}^2,\cdots,\boldsymbol{h}_{K_n,n}^2\right]^{\rm T}\in\mathbb{C}^{K_n\times N_t}$. Then, to serve $K_n$ GTs simultaneously, the UAV first encodes the data symbols for active GTs with a normalized precoding matrix $\boldsymbol{W}_n\in\mathbb{C}^{N_t\times K_n}$. In this paper, we adopt the zero-forcing (ZF) precoder as it can obtain a near-optimal solution at a low complexity. Denote the signal vector for $K_n$ GTs by $\boldsymbol{s}_n\in\mathbb{C}^{K_n\times 1}$, which satisfies $\mathbb{E}\left[\boldsymbol{s}_n\boldsymbol{s}_n^H\right]=P\boldsymbol{I}_{K_n}$. Thus, the received signal at the active GTs in the $n$-th time step can be written by
	\begin{equation}
	\boldsymbol{y}_n=\boldsymbol{H}_n\boldsymbol{W}_n\boldsymbol{s}_n+\boldsymbol{q},
	\end{equation}
	where the $k$-th element of $\boldsymbol{y}_n$ is the received signal for the $k$-th GT and $\boldsymbol{q}\sim \mathcal{CN}\left(\boldsymbol{0},\sigma^2\boldsymbol{{\rm I}}_{K_n}\right)$ is the AWGN vector. Here, we assume that the downlink CSI is perfectly obtained for the UAV by the channel reciprocity of TDD system. For ZF precoding, the precoding matrix $\boldsymbol{W}_n$ can be written by
	\begin{equation}
	\boldsymbol{W}_n=\xi\boldsymbol{H}_n^\dagger,
	\end{equation}
	where $\boldsymbol{H}_n^\dagger=\boldsymbol{H}_n^H\left(\boldsymbol{H}_n\boldsymbol{H}_n^H\right)^{-1}$ and $\xi$ is a constant to meet the total transmitted power constraint after precoding, which can be expressed as
	\begin{equation}
	\xi=\sqrt{\frac{K_n}{{\rm Tr}\left\{\boldsymbol{H}_n^\dagger\left(\boldsymbol{H}_n^\dagger\right)^H\right\}}}.
	\end{equation}
	Consequently, with the ZF precoding, the transmission SNR for the $k$-th GT can be expressed as
	\begin{equation}
	\rho_{k,n}^2=\frac{P{\lVert \boldsymbol{h}_{k,n}\boldsymbol{w}_{k,n} \rVert}^2}{\sigma^2},k\in\mathcal{K}_n.
	\end{equation}
	
	The transmission rate between the UAV and the $k$-th GT can be expressed as
	\begin{equation}
	R_{k,n}=W\log_2\left(1+\rho_{k,n}^2\right),k\in\mathcal{K}_n,
	\end{equation}
	where $W$ is the transmission bandwidth of the UAV. Thus, the hovering time of UAV in the $n$-th time step, which equals to the maximum transmission data duration from the $\mathcal{K}_n$ GTs, can be expressed as
	\begin{equation}
	\delta_{{\rm ht},n}= \max_{k\in\mathcal{K}_n}\left\{ \frac{D_k}{R_{k,n}}\right\},\label{eq-ht}
	\end{equation}
	where $D_k$ denotes the information file size to be received by the $k$-th GT. The completion criterion of the data transmission mission is that all GTs has been served, which can be expressed as
	\begin{equation}
	\sum_{k=1}^K c_{k,N}= K.\label{eq-comp}
	\end{equation}

	Thus, the problem to minimize the mission completion time via trajectory optimization can be formulated as
	\begin{equation}
	\begin{array}{cll}
	\displaystyle\mathop{\mathrm{minimize}}\limits_{\left\{\upsilon_n,\phi_{n}, \theta_{n}\right\},N}\quad & \sum_{n=0}^{N}\delta_n \\ 
	{\rm s.t.} \quad &c_{k,n}=\min\left\{\sum_{i=0}^n \tilde{b}_{k,i}, 1\right\}, \forall n,k,\\
	& \sum_{k=1}^K c_{k,N}= K,\\
	& 0 \le {\upsilon_n} \le \upsilon_{\rm max}, \forall n,\\
	& 0 \le \phi_n \le \pi, \forall n,\\
	& 0 < \theta_n \le 2\pi, \forall n,\\
	& 0 \le x_n \le D, \forall n, \\
	& 0 \le y_n \le D, \forall n, \\
	& z_{\min} \le z_n \le z_{\max}, \forall n, \\
	\end{array}
	\label{Eq8}
	\end{equation}
	where $z_{\min}$ and $z_{\max}$ are the altitude constraints of the UAV. It is noteworthy that the above optimization problem is a mixed-integer non-convex problem, which is known to be NP-hard. Moreover, in the considered scenario, the large-scale fading and small-scale fading depend on the instantaneous locations of the UAV and GTs as well as the surrounding buildings, which makes it to be unrealistic to obtain a closed-form solution. Therefore, it is intractable to solve the above problem by traditional optimization methods like \cite{multi-Rate,multi-efficient}. 
	
	\section{Proposed DRL-Based TDCTM Scheme}\label{III}
	In this section, we reformulate the original problem as a MDP structure and propose the DRL-TDCTM algorithm for UAV trajectory optimization, which aims at minimizing the mission completion time. 
	\subsection{Preliminaries}\label{TD3}
	Reinforcement learning (RL) considers the paradigm of an agent interacting with its environment with the aim of learning reward-maximizing policy\cite{Bible}. Specifically, RL can be used to address a MDP problem with 4-tuple $\left \langle \mathcal S, \mathcal A, \mathcal P, \mathcal R \right \rangle$, where $\mathcal S$ is the state space, $\mathcal A$ is the action space, $\mathcal P$ is the state transition probability, and $\mathcal R$ is the reward function. At each discrete time step $n$, with a given state $s\in \mathcal{S}$, the agent selects action $a\in \mathcal{A}$ with respect to its policy $\pi$, and receives a reward $r$. The return is defined as $R_n=\sum_{i=n}^{N}\gamma^{i-n}r\left(s_i,a_i\right)$, where $\gamma$ is a discount factor determining the priority of short-term rewards.
	
	DRL can be considered as the ``deep" version of RL, which uses multiple DNNs as the approximator of the Q-value function $Q\left(s,a\right)=\mathbb{E}\left[R_n|s,a\right]$. Here $Q\left(s,a\right)$ is the expected return when performing action $a$ in state $s$. In DDPG algorithm\cite{DDPG}, the Q-value approximator $Q_{\theta}\left(s,a\right)$ with parameters $\theta$ can be updated by minimizing the following loss function
	\begin{equation}
	L\left(\theta\right)=\mathbb{E}\left[\left(y-Q_{\theta}\left(s,a\right)\right)^2\right],
	\label{critic-loss}
	\end{equation}
	where $y$ is the target value, which can be estimated by
	\begin{equation}
	y=r+\gamma Q_{\theta^\prime}\left(s^{\prime},a^{\prime}\right),a^{\prime}\sim \pi_{\phi^{\prime}}\left(s^{\prime}\right),
	\end{equation}
	where $s^{\prime}$ is the next state, $a^{\prime}$ is an action selected from a target actor network $\pi_{\phi^{\prime}}$, and $Q_{\theta^\prime}$ is a target network to maintain a fixed objective $y$ over multiple updates. The policy can be updated through the deterministic policy gradient algorithm, which is given by
	\begin{equation}
	{\nabla }_{\phi } {J(\phi)}=\mathbb {E}\left[{\nabla }_{ {a}}  {Q}_{ {\theta}}(s,a) {\vert }_{ {a=}\pi _{\phi } {(s)}}{ {\nabla }}_{\phi }\pi _{ {\phi }}(s)\right].
	\label{actor-loss}
	\end{equation}
	
	As a realization of the celebrated actor-critic algorithm, DDPG can deal with a continuous control problem. Thus, we choose to use it as the starting point for UAV trajectory design with minimum mission completion time.

	\subsection{MDP Formulation}
	Based on optimization problem formulated in Section \uppercase\expandafter{\romannumeral2}-B, we reformulate the original problem of UAV trajectory design with minimum mission completion time as an MDP structure so that DRL algorithm can be applied. In DRL-TDCTM, the UAV is treated as an agent. During the training process, the agent regularly collects the current state information of the environment, then selects a better strategy to control the flight path based on the historical states and rewards. Therefore, we define the state, action, and reward for UAV trajectory design problem as follows.

	1) State $s_n$, $\forall\, n$:
	 $s_n=[b_{1,n},\cdots, b_{K,n}; c_{1,n},\cdots, c_{K,n}; x_n, \\y_n, z_n; \zeta_n]$  is the complete representation of the $n$-th state, which has a cardinality equal to $2K+4$. In state $s_n$, both $b_{k,n}$ and $c_{k,n}$, which have been defined in the Subsection \uppercase\expandafter{\romannumeral2-B}, reflect the data transmission situation of the $k$-th GT; $[x_n,y_n,z_n]$ represents the UAV's 3D position; $\zeta_n$ denotes the merged information between environment and UAV agent during the mission, which can be regarded as an additional information to enhance the decision efficiency and also serves as a reference for reward design. We assume that each GT contains some pheromones, which can be transferred to the UAV. At the same time, pheromones on the UAV will evaporate continuously and more pheromones will evaporate when the UAV's movement violates the boundary. Specifically, $\zeta_n$ can be expressed as 
	\begin{equation}
	\zeta_n = \zeta_{n-1} + K_n\cdot \kappa_{\rm cov}-\kappa_{\rm dis}-P_{\rm ob},
	\end{equation}
	where $\zeta_{n-1}$ is the remaining pheromone in the $\left(n-1\right)$-th time step, $\kappa_{\rm cov}$ is a positive constant that is used to express the captured pheromone per GT, $\kappa_{\rm dis}$ is a positive constant expressing the lost pheromone, and $P_{\rm ob}$ is a penalty when an action causes the boundary violation of the UAV.
	
	2) Action $a_n$, $\forall\, n$: 
	The action is defined as $a_n=[\upsilon_n,\phi_n,\theta_n]$. Since all action variables take continuous values, the UAV's trajectory optimization is a continuous control problem.
	
	3) Reward $r_n$, $\forall\, n$:
	For the above data transmission mission, the UAV agent can not obtain a positive reward until it completes the data transmission for all GTs within the specified time step, i.e., there is no reward in the intermediate process. Furthermore, at the beginning of training, the agent's strategy is random and the reward acquisition needs a series of complex operations. Therefore, the data transmission mission is a sparse rewards problem\cite{Bible}, which, however, may lead to the slow progress over iterations and even non-convergence of RL algorithm. To overcome this issue, we propose a reward shaping mechanism, which can transform the original sparse rewards into dense rewards. Specifically, the reward design is defined as
	\begin{equation} r_{n}=
	\begin{cases}
	r_{\rm tanh}\left(\zeta_n\right)+N_{\rm re}, &{\rm if}\, \sum_{k=1}^K c_{k,n}=K, \\
	r_{\rm tanh}\left(\zeta_n\right), &{\rm otherwise},
	\end{cases}
	\label{Eq9}
	\end{equation}
	where $r_{\rm tanh}\left(\zeta_n\right)=\frac{2}{1+{\rm exp}\left(-\zeta_n/\left(K\cdot \kappa_{\rm cov}\right)\right)}-1$ is a shaped reward function of the pheromone $\zeta_n$. And $r_{\rm tanh}\left(\cdot\right)$ approximates ${\rm tanh}\left(\cdot\right)$ function, but the gradient is smoother than the latter. Due to the dynamic change of pheromone $\zeta_n$, the UAV agent can obtain dense rewards within the exploration stage. Furthermore, the gradient information of reward function can accelerate the convergence of the algorithm. Besides, the UAV would obtain a remaining time reward $N_{\rm re}=N_{\max}-n$ at the mission completion time step, which thus encourages the UAV to complete the data transmission mission as soon as possible.
	
	Combining the DDPG method with the above designs, the DRL-TDCTM is summarized in Algorithm 1.
	
	\begin{algorithm}[tp!] \algsetup{linenosize=\tiny} \scriptsize 
		\caption{DRL-TDCTM}
		\label{alg 1}
		\begin{algorithmic}[1] 
			\STATE Randomly initialize critic network $Q\left(s, a|\theta^Q\right)$ and actor network $\pi\left(s|\theta^\mu\right)$ with weights $\theta^Q$ and $\theta^\mu$
			\STATE Initialize target networks $Q^{\prime}$ and $\mu^{\prime}$ with weights ${\theta^{Q^{\prime}}} \leftarrow \theta^Q$, ${\theta^{\mu^{\prime}}} \leftarrow \theta^\mu$
			\STATE Initialize experience replay buffer $R$
			\FOR{episode $=0$ to $M$} 
			\STATE Initialize the environment, receive an initial state $s_0$, and set $n=0$
			\REPEAT 
			\STATE Select an action $a_n={\pi\left(s_n|\theta^\mu \right)}+\sigma\epsilon$, where $\epsilon$ is a Gaussian noise and $\sigma$ is a decay constant, and observe a reward $r_n=r_{\rm tanh}\left(\zeta_n\right)$ and a new state $s_{n+1}$
			\IF {the UAV flies over the border}
			\STATE $\zeta_n=\zeta_n-P_{\rm ob}$, where $P_{\rm ob}$ is a given penalty. Meanwhile, the movement of the UAV is canceled and update $r_n$, $s_{n+1}$ accordingly
			\ENDIF
			\IF{the UAV completes the data transmission task, i.e., $\sum_{k=1}^K c_{k,n}=K$} 
			\STATE $r_n=r_n+N_{\rm re}$, and the episode is terminated in advance
			\ENDIF
			\STATE Store the transition $\left(s_n, a_n, r_n, s_{n+1}\right)$ in $R$
			\IF{$R>2,000$}
			\STATE Sample a random mini-batch of $B$ transitions from $R$
			\STATE Update critic by minimizing the loss (\ref{critic-loss})
			\STATE Update the actor policy using the sample gradient (\ref{actor-loss})
			\STATE Update the target networks:
			\STATE $\theta^{Q^{\prime}}=\tau\theta^Q+(1-\tau)\theta^{Q^{\prime}}$
			\STATE  $\theta^{\mu^{\prime}}=\tau\theta^\mu+(1-\tau)\theta^{\mu^{\prime}}$
			\ENDIF
			\STATE Update $n\leftarrow n+1$
			\UNTIL{$n=N_{\max}$ or $\sum_{k=1}^K c_{k,n}=K$}
			\ENDFOR
		\end{algorithmic} 
	\end{algorithm}

	\section{Simulation Results}\label{IV}
	In this section, numerical results are conducted to evaluate the performance of the proposed DRL-TDCTM algorithm.
	\subsection{Simulation Settings}
\begin{figure}[htbp]
		\centering
		\subfloat[]{
			\label{fig:2d}
			\includegraphics[width=.45\columnwidth,keepaspectratio]{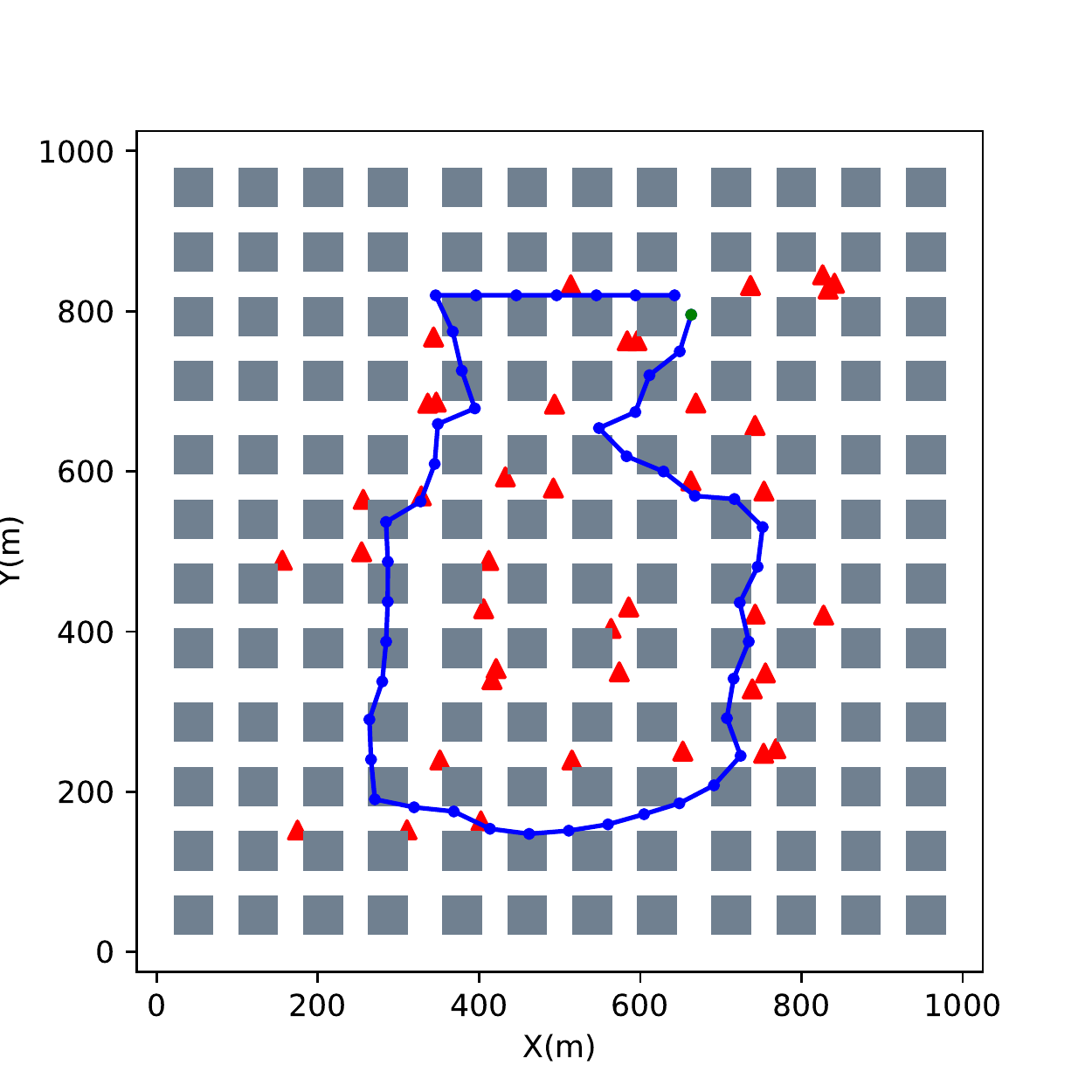}
		}
		\subfloat[]{
			\label{fig:3d}
			\includegraphics[width=.55\columnwidth,keepaspectratio]{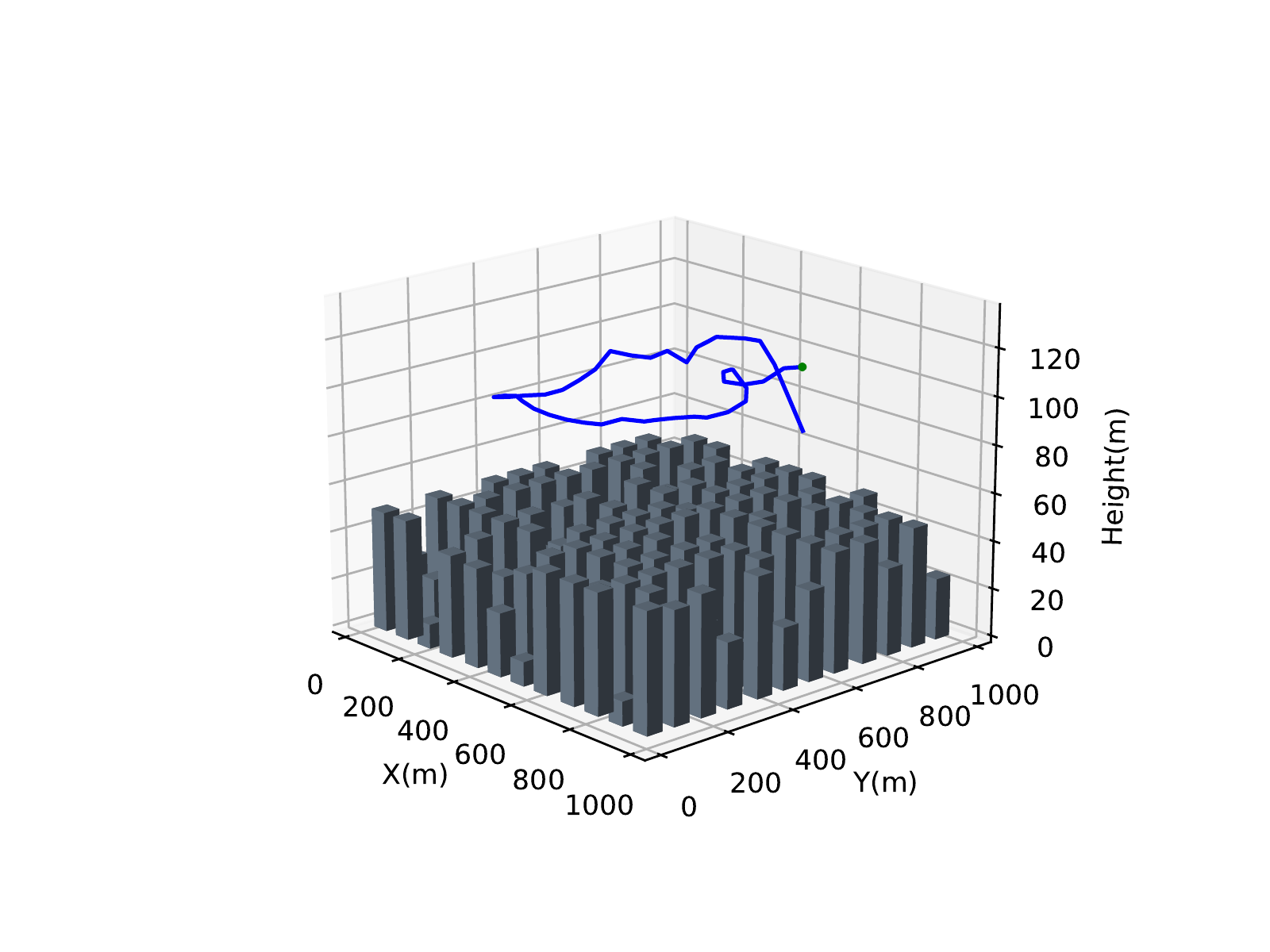}
		}
		\captionsetup{font={footnotesize}, name={Fig.},labelsep=period}
		\caption{UAV's 2D and 3D flight trajectories according to the proposed DRL-TDCTM algorithm, where 40 GTs are considered.}
		\label{fig:view}
		\vspace{-6mm}
	\end{figure}

	As shown in Fig. \ref{fig:view}, we consider an urban area of size $1,000\times1,000$ $\rm{m}^2$ with the dense and high-rise buildings that are generated by one realization of the statistical model in\cite{ITU} with parameters $\alpha=0.3$, $\beta=144$ buildings/km$^2$, and $\lambda=50$ m. To ensure the practicality, the height of building is clipped to $h\in[10, 50]$ m.
	
	We assume that the number of antennas at the UAV is $N_t=12$, the transmit power of the UAV during the broadcast stage is $P=10$ dBm, the noise power is $\sigma^2=-75$ dBm, the SNR threshold is $\rho_{\rm th}=0$ dB satisfying the basic data transmission requirements, the propagation losses are $\eta_{\rm LoS}=0.1$ dB and $\eta_{\rm NLoS}=21$ dB\cite{LAP}, the Rician factor is $G=15$ dB, the information file size to be received by the $k$-th GT is $D_{k}=20$ Mbits, and the transmission bandwidth is $W=5$ MHz. The average flight speed of UAV is assumed to be ${\upsilon_n}\in[0,20]$ m/s, the flight time per step is $\delta_{\rm ft}=2.5 $ s, the hovering time of UAV can be computed by (\ref{eq-ht}), and the altitude constraints of UAV are $z_{\min}=75$ m and $z_{\max}=125$ m. The parameter of pheromone designed is $\kappa_{\rm cov}=10$, and when the UAV is at transmission stage, $\kappa_{\rm dis}=\delta_{{\rm ht},n}$, otherwise, $\kappa_{\rm dis}=2$. As for Algorithm 1, all the actor and critic networks are constructed by a 2-layer fully-connected feedforward neural network with 200 neurons. To encourage the UAV to explore the environment, we add a Gaussian distributed noise $\epsilon\sim \mathcal{N}(0,0.36)$ with a decay rate $\sigma=0.999$ into the action during the training phase. Besides, the maximum number of episodes is $M=8000$, the capacity of the experience replay buffer is $R=1.25\times10^5$, the target network soft-update rate is $\tau=0.005$, the discount factor is $\gamma=0.99$, the mini-batch size is $B=256$, and the maximum time step per episode is $N_{\max}=200$.
	
	Then, we compare DRL-TDCTM with two conventional non-learning based baseline methods.
	
	\begin{itemize}
		\item{Scan strategy: The UAV flies according to a preset path which is a rectangular strip track and it starts from the lower left corner of the area and ends at the upper left corner. Note that such a trajectory design ensures that all locations within the target region are covered by the UAV.}
		\item{ACO-based approach: Taking each GT as a node, it fixes the initial position of the UAV and exploit the ant colony optimization (ACO) algorithm\cite{ant} to solve the shortest path for completing the routing of each node from the determined starting point.}
		
	\end{itemize}
	
	\subsection{Result and Analysis}\label{B}
	To verify the effectiveness of our proposed algorithm, we use the trained model for testing. In each simulation realization, the UAV's initial position is randomly generated. We execute 25 mutually independent realizations in total, whose outputs are averaged to obtain the final results. 
	\begin{figure}[htbp]
		\centering
		\subfloat[]{
			\label{fig:ACT}
			\includegraphics[width=.5\columnwidth,keepaspectratio]{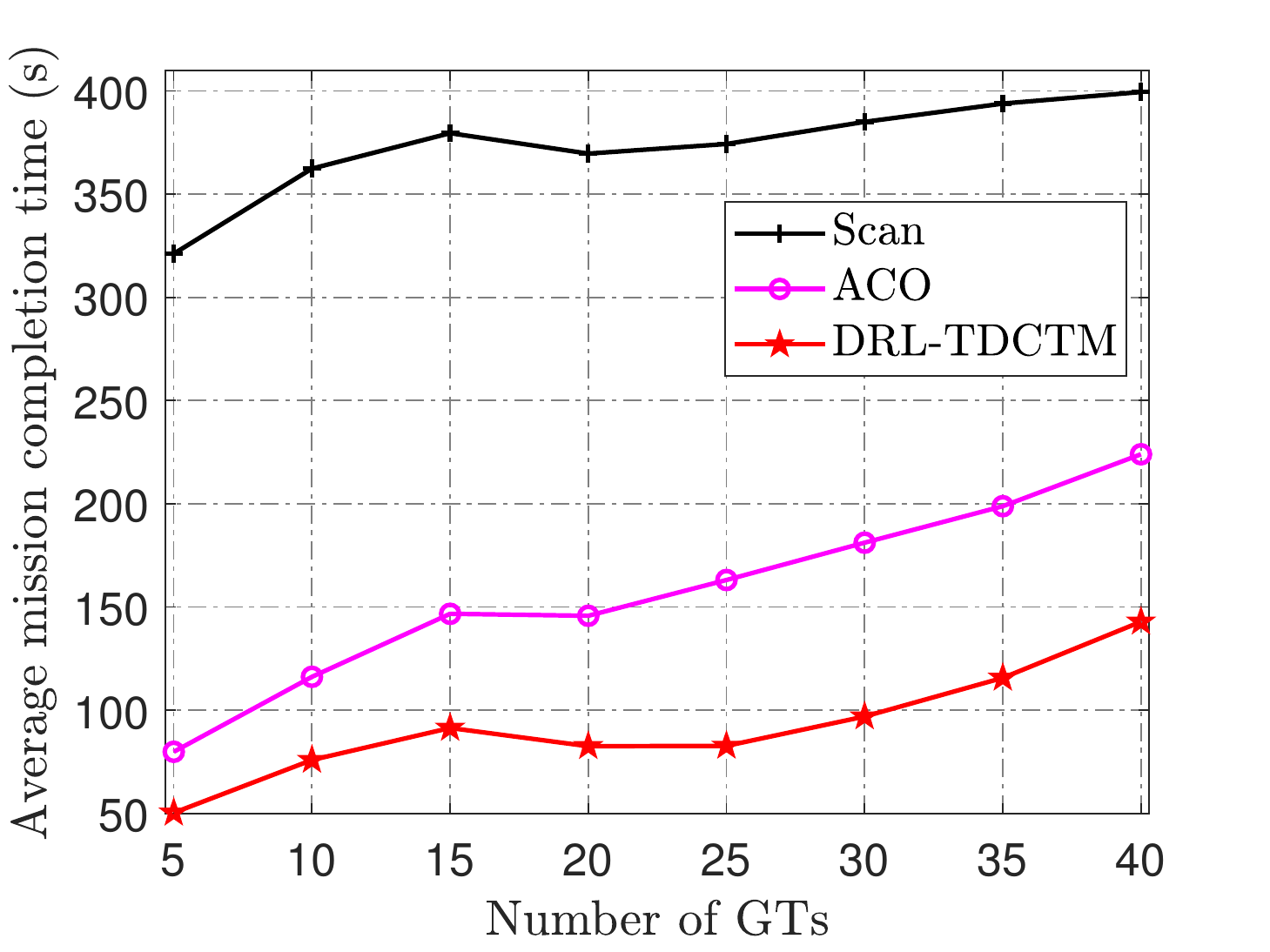}
		}
		\subfloat[]{
			\label{fig:train}
			\includegraphics[width=.5\columnwidth, keepaspectratio]{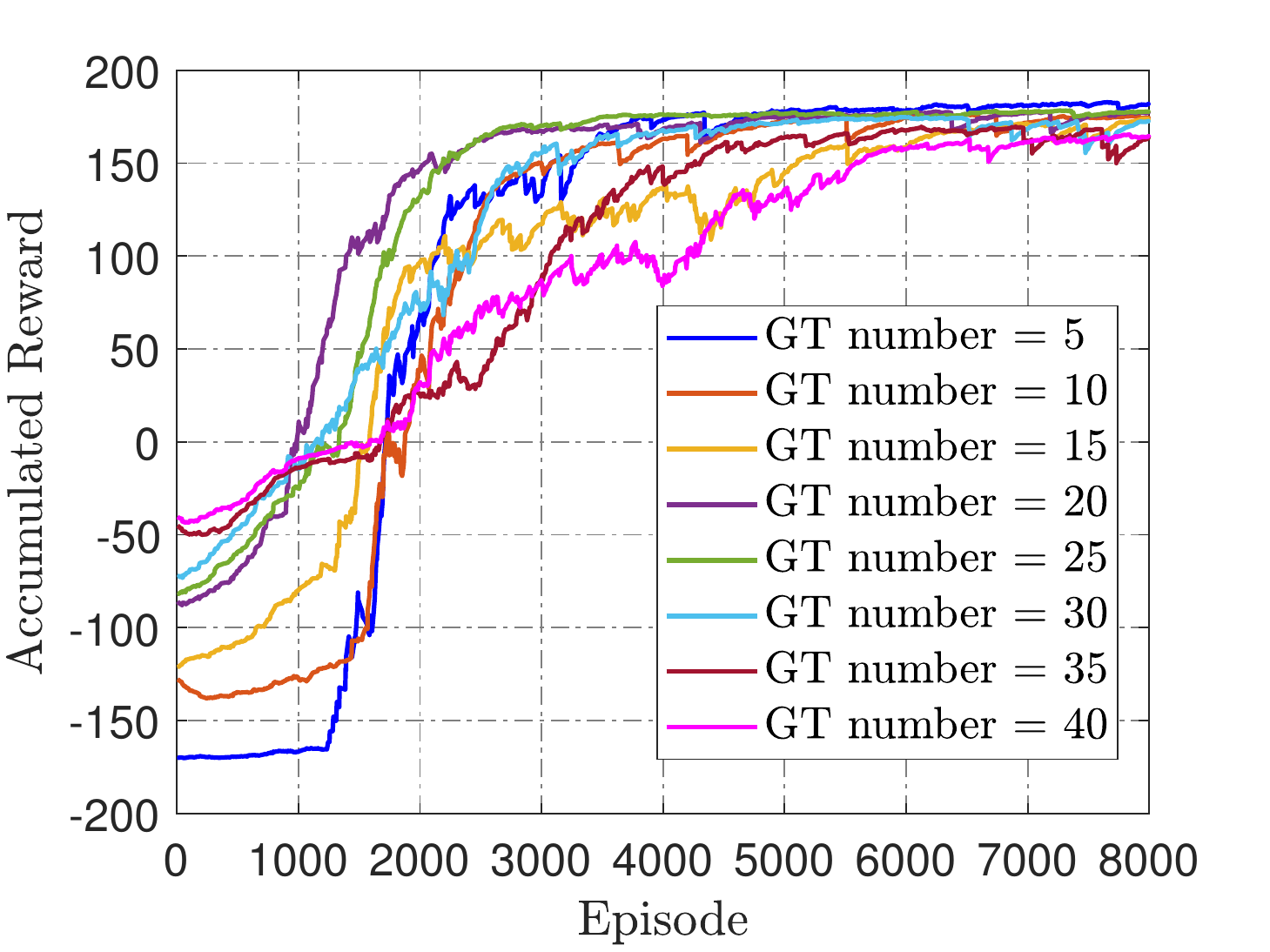}
		}
		\captionsetup{font={footnotesize}, name={Fig.},labelsep=period}
		\caption{The impact of the number of GTs on (a) average mission completion time and (b) convergence performance (i.e., accumulated reward versus episode).}
		\label{fig:perform}
		\vspace{-6mm}
	\end{figure}
	
	In Fig. \ref{fig:view}, the UAV's trajectory is plotted under the case of 40 GTs, where the red triangles represent the served GTs and the blue curve represents the UAV's trajectory. We can observe that the UAV can complete the data transmission mission for all GTs. In such a dense urban environment, buildings are more likely to block the LoS links between the aerial UAV and the terrestrial GTs. Then as the learning process progresses, once the UAV discovers the blockages of LoS links, it would adopt appropriate cruising direction to reestablish the G2A LoS link as soon as possible. Besides, the UAV agent will adaptively adjust its altitude to make a trade-off between the impacts of the LoS probability and the large-scale fading. This fact shows that DRL-TDCTM algorithm can pilot the UAV to sense and learn the external environment. Therefore, it can learn to obtain an approximately optimal strategy for this practical problem with minimum information exchange between the UAV and the environment.
	
	In Fig. \ref{fig:ACT}, we compare the average mission completion time of different methods versus different numbers of GTs. We can observe that the average mission completion time of the proposed DRL-TDCTM algorithm outperforms that of conventional schemes. For 25 GTs, DRL-TDCTM algorithm saves 80.3s compared with the ACO algorithm, and 291.5s compared with the Scan strategy. For the Scan strategy, although the UAV can guarantee to serve all GTs, the exceedingly long mission completion time is intolerable. For the ACO algorithm, although it addresses the shortest route problem from the UAV to each GT, it does not exploit the sensing ability of the UAV, thus there is still a lot of redundancy in flight trajectory. In contrast, the DRL-TDCTM algorithm can sufficiently and adaptively learn how to adjust the exploration strategy. Moreover, these baseline methods can only design the UAV's 2D trajectory, while the proposed method can design the 3D trajectory, which has a higher degree-of-freedom. Therefore, the DRL-TDCTM algorithm can take the minimum time to complete the data transmission task, while ensuring each GT can be served.
	
	Fig. \ref{fig:train} shows the accumulated reward per episode in the training stage under different numbers of GTs. We observe that the accumulated reward shows an upward trend with the increase of the training episodes. After training around 6,000 episodes, the accumulated reward gradually becomes smooth and stable. Besides, the proposed DRL-TDCTM algorithm has the similar convergence performance in the cases of different numbers of GTs. Hence, the proposed scheme is capable of achieving the good convergence and robustness.

	\vspace{-1.5mm}
	\section{Conclusion}
	In this paper, we investigate a multi-user downlink MISO UAV communication system, where a multi-antenna UAV is employed to serve multiple single-antenna GTs. Specifically, we have proposed a DRL-based efficient 3D trajectory design, DRL-TDCTM, to minimize the transmission mission completion time in a 3D urban environment. In particular, we set an additional information, i.e., the merged pheromone, to enhance the decision efficiency. By taking the service status of IoT nodes, the UAV's position, and the merged pheromone as input, the DRL-TDCTM algorithm can continuously and adaptively learn how to adjust the UAV's movement strategy for minimizing the completion time under the constraints in flight movement and throughput. Numerical results show a significant performance gain of the DRL-TDCTM algorithm over the existing baseline methods.


\end{document}